\documentclass[12pt]{iopart}
\usepackage{amsfonts}
\usepackage{amssymb}
\usepackage{graphicx}
\usepackage{setspace}
\usepackage{CJK}
\usepackage{moresize}
\usepackage[utf8]{inputenc}
\usepackage[english]{babel}

\begin{document}

\title{A connection between linearized Gauss-Bonnet gravity and classical electrodynamics}
\date{\today}

\linespread{1.25}

\author{Mark Robert Baker$^{1,2}$ and Sergei Kuzmin$^{1,3}$ }

\address{$^1$ Department of Physics and Astronomy, University of Western Ontario, London, ON, N6A 3K7, Canada}

\address{$^2$ The Rotman Institute of Philosophy, University of Western Ontario, London, ON, N6A 5B7, Canada}

\address{$^3$ Department of Economics, Business and Mathematics, King's University College, London, ON, N6A 2M3, Canada}

\ead{mbaker66@uwo.ca and skuzmin@uwo.ca}

\vspace{10pt}

\begin{indented}
\item[]31 Oct 2018
\end{indented}

\begin{abstract}
A connection between linearized Gauss-Bonnet gravity and classical electrodynamics is found by developing a procedure which can be used to derive completely gauge invariant models. The procedure involves building the most general Lagrangian for a particular order of derivatives ($N$) and rank of tensor potential ($M$), then solving such that the model is completely gauge invariant (the Lagrangian density, equation of motion and energy-momentum tensor are all gauge invariant). In the case of $N = 1$ order of derivatives and $M = 1$ rank of tensor potential, electrodynamics is uniquely derived from the procedure. In the case of $N = 2$ order of derivatives and $M = 2$ rank of symmetric tensor potential, linearized Gauss-Bonnet gravity is uniquely derived from the procedure. The natural outcome of the models for classical electrodynamics and linearized Gauss-Bonnet gravity from a common set of rules provides an interesting connection between two well explored physical models.
\end{abstract}

\maketitle

\section{Motivation}

\normalsize

Gauge invariance is a common characteristic among field theories of fundamental interactions. Electrodynamics has a unique property with respect to gauge invariance that will be called {\it{complete gauge invariance}} in this article. Complete gauge invariance occurs when the Lagrangian density, equation of motion and energy-momentum tensor are all gauge invariant \cite{burgess2002}. Complete gauge invariance of the model is possible only when the Lagrangian density is exactly gauge invariant. Consider the Noether identity for a general potential $\Phi_A$ \cite{noether1918,kosmann2011noether,jackiw1994},

\begin{equation}
\fl
\eqalign{ \left( \frac{\partial \mathcal{L}}{\partial \Phi_A}
 - \partial_\mu \frac{\partial \mathcal{L}}{\partial (\partial_\mu \Phi_A)} 
 + \partial_\mu \partial_\omega \frac{\partial \mathcal{L}}{\partial (\partial_\mu \partial_\omega \Phi_A)} + \dots \ \right) \delta \Phi_A
 \\
+ \partial_\mu \left(  \eta^{\mu\nu} \mathcal{L} \delta x_\nu
+ \frac{\partial \mathcal{L}}{\partial (\partial_\mu \Phi_A)} \delta \Phi_A 
+ \frac{\partial \mathcal{L}}{\partial (\partial_\mu \partial_\omega \Phi_A)} \partial_\omega \delta \Phi_A
- \left[ \partial_\omega \frac{\partial \mathcal{L}}{\partial (\partial_\mu \partial_\omega \Phi_A)} \right] \delta \Phi_A
+ ... \ \right) = 0 \ . }
   \label{genenergy}
\end{equation}

This identity is the result of invariance of the action under simultaneous change of coordinates and fields. For a given coordinate change $\delta x_\nu$ and corresponding change of fields $\delta \Phi_A$ of the Lagrangian density, a conservation law will follow from the particular form of $\delta \Phi_A$, such as in the case of Lorentz translation the energy-momentum tensor is derived. All conservation laws follow from the expression under the total divergence. The conservation law depends explicitly on the Lagrangian density which restricts gauge invariant energy-momentum tensors to those which have explicitly gauge invariant Lagrangian densities. In electrodynamics, $\mathcal{L} = - \frac{1}{4} F_{\alpha\beta} F^{\alpha\beta}$ is exactly gauge invariant, so this is not a problem. 

The equation of motion for the spin-2 model \cite{fierz1939} is gauge invariant under the transformation $h_{\mu\nu}' = h_{\mu\nu} + \partial_\mu \xi_\nu + \partial_\nu \xi_\mu$. For spin-2, the Fierz-Pauli Lagrangian density  ($\mathcal{L} = \frac{1}{4}[\partial_\alpha h_\beta^\beta \partial^\alpha h_\gamma^\gamma - \partial_\alpha h_{\beta\gamma} \partial^\alpha h^{\beta\gamma} + 2 \partial_\alpha h_{\beta\gamma} \partial^\gamma h^{\beta\alpha} - 2 \partial^\alpha h_\beta^\beta \partial^\gamma h_{\gamma\alpha}]$) is not exactly invariant under a gauge transformation \cite{magnano2002,padmanabhan2008}, even after a change of variables \cite{ mckeon1979}. It is only gauge invariant up to the surface term $\delta_g \mathcal{L} =  \partial_\mu [
h^{\nu\gamma} \partial_\nu \partial_\gamma \xi^\mu
- \frac{1}{2} h^{\mu\nu} \partial_\nu \partial_\gamma \xi^\gamma 
+ \frac{1}{2} h \partial^\mu \partial_\gamma \xi^\gamma 
- \frac{1}{2} h \square \xi^\mu 
]$. This is why the energy-momentum tensor is not gauge invariant \cite{magnano2002,dewit1980}. Often this fact is overlooked due to the common priority that only an equation of motion must be found which is gauge invariant. It will be shown in this article why for spin-2 there exists no exactly gauge invariant Lagrangian or gauge invariant conservation law; the canonical energy-momentum tensor depends explicitly on the Lagrangian density \cite{jackiw1994,magnano2002}.

It is clear that exact invariance of the action is a special property, as indicated by electrodynamic theory. The motivation for the current work is as follows: develop a procedure such that an explicitly gauge invariant Lagrangian can be derived. From this procedure, models can be constructed that are invariant under a desired gauge transformation. The power of this procedure is highlighted by a result that was not foreseen by the development of the procedure; the model constructed from a general Lagrangian density which is quadratic in second order derivatives of a  symmetric second rank tensor potential (i.e. $\partial_\alpha \partial_\beta h_{\mu\nu} \partial^\alpha \partial^\beta h^{\mu\nu}$), and invariant under the spin-2 gauge transformation, is uniquely the linearized Gauss-Bonnet gravity model. The Gauss-Bonnet gravity model is a frequent topic in the physics literature \cite{zwiebach1985,escalante2004,charmousis2002,wheeler1986}; a connection between linearized Gauss-Bonnet gravity and classical electrodynamics can give some additional insight into the significance of the Gauss-Bonnet model.

\section{A Procedure For Gauge Invariant Lagrangian Formulation}

To derive a completely gauge invariant model, a procedure was developed that would yield an exactly gauge invariant Lagrangian, equation of motion and energy-momentum tensor. We restrict our attention to Poincar\'e invariant field theories. The procedure involves defining a linear combination of all possible contractions of terms quadratic in derivatives of fields (i.e. for a model built using a vector potential quadratic in first order derivatives, such as $\partial_\mu A_\nu \partial^\mu A^\nu$).  Once the general Lagrangian is constructed, a gauge transformation is applied, and a linear system of equations is obtained. This leads to specific coefficients that will yield a gauge invariant Lagrangian. From here, we obtain a gauge invariant equation of motion, and Noether's theorem is used to derive an energy-momentum tensor \cite{
 noether1918,kosmann2011noether}. The procedure can be applied for any order $N$ of derivatives/ rank $M$ of tensor potential.

For a model built using a vector potential which is quadratic in first order derivatives ($N = 1$, $M = 1$ i.e. $\partial_\mu A_\nu \partial^\mu A^\nu$), this procedure yields exactly the electrodynamic Lagrangian as the unique gauge invariant combination. For spin-2 ($N = 1$, $M = 2$ i.e. $\partial_\mu h_{\nu \gamma} \partial^\mu h^{\nu \gamma} $), this procedure yields only $\mathcal{L} = 0$. It is possible to modify this procedure to consider Lagrangian densities which are invariant up to some surface term (such as the Fierz-Pauli action), but this does not concern completely gauge invariant models. While spin-2 does not have an explicitly gauge invariant action, a symmetric second rank potential $h_{\nu \gamma}$ does in fact belong to a completely gauge invariant model for $N =2$, $M =2$, which requires higher order derivatives in the general Lagrangian ($N =2$, $M =2$ i.e. $\partial_\mu \partial_\nu h_{\alpha\beta} \partial^\mu \partial^\nu h^{\alpha\beta}$). The procedure for $N = 2$, $M = 2$ derives a unique completely gauge invariant model, which is exactly the linearized Gauss-Bonnet model for gravity. 

\subsection{Derivation for $N = 1$, $M = 1$ (classical electrodynamics)}

For a model with $N = 1$ and $M = 1$ (i.e. vector field theory), the Lagrangian is a scalar which is quadratic in first order field derivatives (i.e. $\partial_\mu A_\nu \partial^\mu A^\nu$). To consider all possible combinations,  first write all possible contractions of two indices. From here, write all possible contractions of the next two indices. Since there are only two contracting pairs of indices, this procedure is extremely simple, yielding a linear combination of only 3 possible terms,

\begin{equation}
\mathcal{L} = a \partial_\mu A_\nu \partial^\mu A^\nu + b \partial_\mu A^\mu \partial_\nu A^\nu + c \partial_\mu A_\nu \partial^\nu A^\mu ,  \label{spin1lagrangian}
\end{equation}

where $a, b, c$ are arbitrary coefficients. Imposing a gauge transformation $A_\mu' = A_\mu + \partial_\mu \phi$ yields a Lagrangian density which can be organized by the original terms $(\mathcal{L})$, and transformation terms ($\delta_g \mathcal{L}$), in the form $\mathcal{L}' = \mathcal{L} + \delta_g \mathcal{L}$,

\begin{equation}
\mathcal{L}' = \mathcal{L}
+ 2 (a + c) \partial_\mu A_\nu \partial^\mu \partial^\nu \phi 
 + 2  b \partial_\mu A^\mu \partial_\nu \partial^\nu \phi , 
\end{equation}

where $\mathcal{L}$ is given in (\ref{spin1lagrangian}). A system of equations is derived such that if all of the $\delta_g \mathcal{L}$ cancel, an exactly gauge invariant Lagrangian density will be obtained. In other words, we must solve for $\delta_g \mathcal{L} = 0$. In order to satisfy this condition, we require $a + c = 0$ and $b = 0$, thus $c = -a$. The resulting Lagrangian density can be factored to $\mathcal{L} = \frac{1}{2} a (\partial_\mu A_\nu - \partial_\nu A_\mu)(\partial^\mu A^\nu  -  \partial^\nu A^\mu) $.

What we find is that the Lagrangian density is exactly the form of classical electrodynamics. The arbitary coefficient $a$ allows for the standard coefficient of the electrodynamic Lagrangian density, $a = -\frac{1}{2}$ yields $\mathcal{L} = - \frac{1}{4} F_{\mu\nu}F^{\mu\nu}$. The choice of the standard coefficient is fixed by equation of motion (Maxwell's equations) and the energy-momentum tensor $T^{\mu\nu} = F^{\mu\alpha}F^\nu_\alpha - \frac{1}{4} \eta^{\mu\nu} F_{\alpha\beta} F^{\alpha\beta}$. The electromagnetic energy-momentum tensor and all other conservation laws related to conformal invariance of Maxwell's equations were first derived from Noether's theorem by Bessel-Hagen \cite{besselhagen1921}. This also can be derived from the canonical Noether energy-momentum tensor after implementing the Belinfante improvement \cite{belinfante1940}. The procedure therefore can be used to derive a completely gauge invariant model with no free coefficients, in the case of $N=1$ and $M=1$, classical electrodynamics.

\subsection{Derivation for $N = 1$, $M = 2$ (spin-2)}

For a model with $N = 1$ and $M = 2$ (i.e. tensor field theory based on symmetric $h_{\nu \gamma}$), the Lagrangian is a scalar which is quadratic in first order field derivatives (i.e. $\partial_\mu h_{\nu \gamma} \partial^\mu h^{\nu \gamma} $). Avoiding redundant contractions yields \cite{green2011},

\begin{equation}
\fl
\mathcal{L} = A \partial_\mu h^\mu_\nu \partial^\nu h_\gamma^\gamma
+ B \partial_\mu h^\mu_\nu \partial_\gamma h^{\nu \gamma}
+ C \partial_\mu h_\nu^\nu \partial^\mu h_\gamma^\gamma 
+ D \partial_\mu h_{\nu \gamma} \partial^\mu h^{\nu \gamma} 
+ E \partial_\mu h_{\nu \gamma} \partial^\nu h^{\mu \gamma}  , \label{spin2lag}
\end{equation}

where $A, B, C, D, E$ are arbitrary coefficients. Imposing a spin-2 gauge transformation $h_{\mu\nu}' = h_{\mu\nu} + \partial_\mu \xi_\nu + \partial_\nu \xi_\mu$ yields $\mathcal{L}' = \mathcal{L} + \delta_g \mathcal{L}$, with $\delta_g \mathcal{L}$ which must vanish in order for a gauge invariant expression to be derived. For clarity, common terms are combined and the D'Alembertian operator ($\square = \partial_\mu \partial^\mu$) is introduced. The resulting Lagrangian density is,

\begin{equation}
\fl
\eqalign{
\mathcal{L}' = \mathcal{L} + A \partial^\nu h \square \xi_\nu
+ 2 B \partial_\gamma h^{\nu\gamma} \square \xi_\nu 
+ (A + 4 C) \partial_\mu h \partial^\mu \partial_\gamma \xi^\gamma 
\\
+ (2 A + 2 B) \partial_\mu h^{\mu\nu} \partial_\nu \partial_\gamma \xi^\gamma 
+ (4 D + 2 E) \partial_\mu h_{\nu\gamma} \partial^\mu \partial^\nu \xi^\gamma 
+ 2 E \partial_\mu h_{\nu\gamma} \partial^\nu \partial^\gamma \xi^\mu , } 
\end{equation}

This equation leads to the homogenous linear system which has either a trivial solution, or a non-trivial solution with free parameter(s). The trivial gauge invariant Lagrangian $\mathcal{L} = 0$, is the only gauge invariant expression. This is the expected result because spin-2 is well known to have an action which is gauge invariant only up to a surface term \cite{dewit1980,magnano2002}. We can find this surface term by using integration by parts, leaving us with a total divergence and some remaining terms,

\begin{equation}
\fl
\eqalign{
\mathcal{L}' = \mathcal{L} + \partial_\mu [A h \square \xi^\mu
+ 2 B h^{\nu\mu} \square \xi_\nu 
+ (A + 4 C) h \partial^\mu \partial_\gamma \xi^\gamma 
\\
+ (2 A + 2 B) h^{\mu\nu} \partial_\nu \partial_\gamma \xi^\gamma 
+ (4 D + 2 E) h_{\nu\gamma} \partial^\mu \partial^\nu \xi^\gamma 
+ 2 E h_{\nu\gamma} \partial^\nu \partial^\gamma \xi^\mu ]
\\
- (2 A + 4 C) h \square \partial_\gamma \xi^\gamma 
- (2 A + 2 B + 2 E) h^{\mu\nu} \partial_\mu \partial_\nu \partial_\gamma \xi^\gamma 
- (2 B + 4 D + 2 E) h_{\nu\gamma} \square \partial^\nu \xi^\gamma  . }
\end{equation}

We solve for the coefficients such that the terms not under the total divergence are identically zero. Solving this system of linear equations we have $C = - \frac{1}{2} A$, $D = \frac{1}{2} A$ and $E = - A - B$. For any choice of the free coefficient we have a Lagrangian density which yields the spin-2 equation of motion. If we solve such that the coefficients match up with the conventional coefficient $\frac{1}{2}$ of the spin-2 equation of motion, we must take $A = -\frac{1}{2}$, thus we are left with the solution $A = - \frac{1}{2}$, $B = B$, $C = \frac{1}{4}$, $D = - \frac{1}{4}$ and $E = \frac{1}{2} - B$ with only one free parameter $B$. We are left with a one parameter family of Lagrangian densities which is invariant up to the surface term,

\begin{equation}
\fl
\eqalign{
\mathcal{L}' = \mathcal{L} + \partial_\mu [
h^{\nu\gamma} \partial_\nu \partial_\gamma \xi^\mu
-  h^{\mu\nu} \partial_\nu \partial_\gamma \xi^\gamma 
+ \frac{1}{2} h \partial^\mu \partial_\gamma \xi^\gamma 
- \frac{1}{2} h \square \xi^\mu 
\\
+ 2 B (h^{\mu\nu} \partial_\nu \partial_\gamma \xi^\gamma 
+ h^{\nu\mu} \square \xi_\nu
-  h_{\nu\gamma} \partial^\mu \partial^\nu \xi^\gamma 
 - h_{\nu\gamma} \partial^\nu \partial^\gamma \xi^\mu)
] . }
\end{equation}

The coefficients used to ensure the Lagrangian is invariant up to a surface term from (\ref{spin2lag}),

\begin{equation}
\fl
\mathcal{L} = \frac{1}{4} [- 2 \partial_\mu h^\mu_\nu \partial^\nu h_\gamma^\gamma
+ 4 B \partial_\mu h^\mu_\nu \partial_\gamma h^{\nu \gamma}
+  \partial_\mu h_\nu^\nu \partial^\mu h_\gamma^\gamma 
-  \partial_\mu h_{\nu \gamma} \partial^\mu h^{\nu \gamma} 
+ (2 + 4 B) \partial_\mu h_{\nu \gamma} \partial^\nu h^{\mu \gamma} ]  .
\end{equation}

We note that for particular value of the free parameter $B = 0$, the Fierz-Pauli Lagrangian is recovered identically. Having a Lagrangian which is explicitly gauge invariant is likely related to the ability to construct a field strength tensor, as will be emphasized during the construction of the gauge invariant model in the following section. Absence of a gauge invariant Lagrangian seems to imply an inability to construct a quadratic combination of independently gauge invariant field strength tensors. The notion of the field strength tensor in a physical model has be alluded to as a physics necessity in the past \cite{darrigol2014,landau1951classical,blagojevic2013}; the current work can give some more insight into these observations.

\subsection{Derivation for $N = 2$, $M = 2$ (linearized Gauss-Bonnet)}

For a model with $N = 2$ and $M = 2$, the Lagrangian is a scalar which is quadratic in second order field derivatives (i.e. $\partial_\mu \partial_\nu h_{\alpha\beta} \partial^\mu \partial^\nu h^{\alpha\beta} $). This can be built from the Lagrangian of the form $\mathcal{L} = M^{\mu\nu\alpha\beta\rho\lambda\sigma\gamma} \partial_\mu \partial_\nu h_{\alpha\beta} \partial_\rho \partial_\lambda h_{\sigma\gamma}$, where $ M^{\mu\nu\alpha\beta\rho\lambda\sigma\gamma}$ is all possible permutations of indices of four Minkowski tensors (i.e. $\eta^{\mu\rho} \eta^{\nu\lambda} \eta^{\alpha\sigma} \eta^{\beta\gamma}$). Avoiding terms which are redundant after contraction yields, 

\begin{equation}
\fl
\eqalign{
\mathcal{L} = C_{1} \partial_\mu \partial^\mu h_\nu^{\nu} \partial_\alpha \partial^\alpha h_{\beta}^{\beta} 
+ C_{2} \partial_\mu \partial^\mu h_{\alpha\beta} \partial_\nu \partial^\nu h^{\alpha\beta} 
+ C_{3} \partial_\mu \partial_\nu h^{\mu\nu} \partial_\alpha \partial^\alpha h_{\beta}^{\beta} \\
+ C_{4} \partial_\mu \partial_\nu h_{\alpha}^{\alpha} \partial_\beta \partial^\beta h^{\mu\nu} 
+ C_{5} \partial_\mu \partial_\nu h^{\nu}_{\beta} \partial_\alpha \partial^\alpha h^{\mu\beta} 
+ C_{6} \partial_\mu \partial_\nu h_{\alpha}^{\alpha} \partial^\mu \partial^\nu h_{\beta}^{\beta} 
+ C_{7} \partial_\mu \partial_\nu h_{\alpha}^{\alpha} \partial^\mu \partial_\beta h^{\nu\beta} \\
+ C_{8} \partial_\mu \partial_\nu h^{\mu\nu} \partial_\alpha \partial_\beta h^{\alpha\beta} 
+ C_{9} \partial_\mu \partial_\nu h^{\nu\beta} \partial^\mu \partial_\alpha h^\alpha_\beta 
+ C_{10} \partial_\mu \partial_\nu h^\nu_{\beta} \partial^\beta \partial_\alpha h^{\mu\alpha} \\
+ C_{11} \partial_\mu \partial_\nu h_{\alpha\beta} \partial^\mu \partial^\nu h^{\alpha\beta} 
+ C_{12} \partial_\mu \partial_\nu h_{\alpha\beta} \partial^\mu \partial^\alpha h^{\nu\beta} 
+ C_{13} \partial_\mu \partial_\nu h_{\alpha\beta} \partial^\alpha \partial^\beta h^{\mu\nu} . }
\end{equation}

A spin-2 gauge transformation $h_{\mu\nu}' = h_{\mu\nu} + \partial_\mu \xi_\nu + \partial_\nu \xi_\mu$ is then applied. The $ \delta_g \mathcal{L}$ consists of 10 unique terms which result in a system of 10 linear equations. These equations decouple into 3 independent systems which we will call i), ii) and iii), each of which are solvable with one free parameter. Not a single coefficient $C_n$ is zero in the case of a nontrivial gauge invariant Lagrangian ($\delta_g \mathcal{L} = 0$). Each of the independent system of linear equations can be solved, i) $C_{12} = -2 C_{11}, \ C_{13} = C_{11}$, ii) $ C_{4} = 2 C_{2}, \ C_{5} = -4 C_{2}, \ C_{6} = C_{2}, \ C_{7} = -4 C_{2}, \ C_{9} = 2 C_{2}, \ C_{10} = 2 C_{2}$ and iii) $ C_3 = -2 C_1, \ C_8 = C_1$. These solutions yield the following independently gauge invariant combinations,

\begin{equation}
\fl
\eqalign{
\mathcal{L} = C_{11} (\partial_\mu \partial_\nu h_{\alpha\beta} \partial^\mu \partial^\nu h^{\alpha\beta} - 2 \partial_\mu \partial_\nu h_{\alpha\beta} \partial^\mu \partial^\alpha h^{\nu\beta} + \partial_\mu \partial_\nu h_{\alpha\beta} \partial^\alpha \partial^\beta h^{\mu\nu})
\\
+  C_{2} ( \partial_\mu \partial^\mu h_{\alpha\beta} \partial_\nu \partial^\nu h^{\alpha\beta} + 2 \partial_\mu \partial_\nu h_{\alpha}^{\alpha} \partial_\beta \partial^\beta h^{\mu\nu} 
 - 4 \partial_\mu \partial_\nu h^{\nu}_{\beta} \partial_\alpha \partial^\alpha h^{\mu\beta} 
 \\
 + \partial_\mu \partial_\nu h_{\alpha}^{\alpha} \partial^\mu \partial^\nu h_{\beta}^{\beta} 
 - 4 \partial_\mu \partial_\nu h_{\alpha}^{\alpha} \partial^\mu \partial_\beta h^{\nu\beta} + 2 \partial_\mu \partial_\nu h^{\nu\beta} \partial^\mu \partial_\alpha h^\alpha_\beta + 2 \partial_\mu \partial_\nu h^\nu_{\beta} \partial^\beta \partial_\alpha h^{\mu\alpha} )
\\
+ C_1 (\partial_\mu \partial^\mu h_\nu^{\nu} \partial_\alpha \partial^\alpha h_{\beta}^{\beta} - 2 \partial_\mu \partial_\nu h^{\mu\nu} \partial_\alpha \partial^\alpha h_{\beta}^{\beta} + \partial_\mu \partial_\nu h^{\mu\nu} \partial_\alpha \partial_\beta h^{\alpha\beta}) . }
\end{equation}

The 3 combinations can be factored into contractions of a fourth rank, second rank, and zeroth rank tensor. The motivation is to have each term expressed as the contraction of two field strength tensors. The result of this,

\begin{equation}
\fl
\eqalign{
\mathcal{L} =  \frac{1}{4} C_{11} (\partial_\mu \partial_\alpha h_{\nu\beta} + \partial_\nu \partial_\beta h_{\mu\alpha} - \partial_\mu \partial_\beta h_{\nu\alpha} - \partial_\nu \partial_\alpha h_{\mu\beta} ) ( \partial^\nu \partial^\beta h^{\mu\alpha} + \partial^\mu \partial^\alpha h^{\nu\beta} - \partial^\nu \partial^\alpha h^{\mu\beta} - \partial^\mu \partial^\beta h^{\nu\alpha})
\\
+ C_{2} (\square h_{\mu\nu} + \partial_\mu \partial_\nu h - \partial_\mu \partial_\alpha h^{\alpha}_{\nu} - \partial_\nu \partial_\alpha h^{\alpha}_{\mu})
 (\square h^{\mu\nu} 
+ \partial^\mu \partial^\nu h 
-  \partial^\mu \partial^\alpha h_{\alpha}^{\nu} 
- \partial^\nu \partial^\alpha h_{\alpha}^{\mu})
\\
+ C_{1} (\square h - \partial_\mu \partial_\nu h^{\mu\nu})(\square h -  \partial_\alpha \partial_\beta h^{\alpha\beta}) , }
\end{equation}

shows 3 expressions which are familiar to Riemannian geometry. They are the linearized terms of the Riemann tensor $R^{\mu\nu\alpha\beta} = \frac{1}{2} (  \partial^\mu \partial^\beta h^{\nu\alpha} + \partial^\nu \partial^\alpha h^{\mu\beta} -\partial^\mu \partial^\alpha h^{\nu\beta} - \partial^\nu \partial^\beta h^{\mu\alpha})$, and Ricci tensor $R^{\nu\beta} = \eta_{\mu\alpha} R^{\mu\nu\alpha\beta} = \frac{1}{2} (  
 \partial^\beta \partial^\alpha h_{\alpha}^{\nu}
+\partial^\nu \partial^\alpha h_{\alpha}^{\beta} 
-\square h^{\nu\beta} 
- \partial^\nu \partial^\beta h 
)$, and Ricci scalar $R = \eta_{\nu\beta} R^{\nu\beta} =  \partial_\mu \partial_\nu h^{\mu\nu} - \square h$. Using these linearized Riemann/Ricci tensors, and rewriting the free coefficients as $\tilde{a}$, $\tilde{b}$ and $\tilde{c}$ we have,

\begin{equation}
\mathcal{L} = \tilde{a} R_{\mu\nu\alpha\beta} R^{\mu\nu\alpha\beta} + \tilde{b} R_{\mu\nu} R^{\mu\nu} + \tilde{c} R^2 ,  \label{gaugeL}
\end{equation}

which allows for an infinite number of possible models to be developed. The goal of this work is specific, to find unique combinations which lead to completely gauge invariant models. This condition will now be used to specify possible Lagrangian densities, by use of Noether's theorem. Referring again to (\ref{genenergy}) we have the conservation law for $N=2,M=2$,

  \begin{equation}
   \partial_\omega \left[  \eta^{\omega\nu} \mathcal{L} \delta x_\nu +
  \frac{\partial \mathcal{L}}{\partial (\partial_\omega \partial_\lambda h_{\rho\sigma})} \partial_\lambda \delta h_{\rho\sigma} 
 - \left( \partial_\lambda \frac{\partial \mathcal{L}}{\partial (\partial_\omega \partial_\lambda h_{\rho\sigma})} \right) \delta h_{\rho\sigma} \right] = 0. \label{ourgenen}
\end{equation}

The first term is gauge invariant because the Lagrangian density is explicitly gauge invariant. The second term can possibly be gauge invariant depending on transformation of the fields $\partial_\lambda \delta h_{\rho\sigma} $ since there will be second order derivatives of $ h_{\rho\sigma}$. From (\ref{gaugeL}) we calculate,

\begin{equation}
\fl
\eqalign{
\frac{\partial \mathcal{L}}{\partial (\partial_\omega \partial_\lambda h_{\rho\sigma})} =  2 \tilde{a} [R^{\rho\omega\lambda\sigma} +  R^{\lambda\rho\sigma\omega}]    
\\
 + \tilde{b} [ - \eta^{\omega \lambda} R^{\rho\sigma} - \eta^{\rho \sigma} R^{\omega\lambda} + \frac{1}{2} ( \eta^{\lambda \sigma } R^{\rho\omega} +  \eta^{\lambda \rho} R^{\sigma\omega} +  \eta^{\omega \sigma} R^{\rho\lambda} +  \eta^{\omega \rho} R^{\sigma\lambda} ) ]
\\
+   2 \tilde{c} R [-\eta^{\omega\lambda} \eta^{\rho\sigma} + \frac{1}{2} (\eta^{\omega\rho} \eta^{\lambda\sigma} + \eta^{\omega\sigma} \eta^{\lambda\rho} )] . } \label{genderiv}
\end{equation}

Two identities can be used in the following calculations which are equivalent to the Bianchi identities: $\partial_\omega R^{\lambda\rho\omega\sigma} = \partial^\lambda R^{\rho\sigma} - \partial^\rho R^{\lambda\sigma}$ and $\partial^\rho R = 2 \partial_\omega R^{\omega\rho}$. These identities allow for all of the tensors to be expressed as derivatives of the Ricci tensor. Using (\ref{genderiv}), we have for the second term in (\ref{ourgenen}),

 \begin{equation}
 \fl
\eqalign{
 \frac{\partial \mathcal{L}}{\partial (\partial_\omega \partial_\lambda h_{\rho\sigma})} \partial_\lambda \delta h_{\rho\sigma} =
 2 \tilde{a} R^{\omega\rho\lambda\sigma} (\partial_\sigma \delta h_{\lambda\rho} - \partial_\lambda \delta h_{\sigma\rho})
 + \tilde{b} R^{\rho\sigma} (\partial_\rho \delta h^\omega_\sigma - \partial^\omega \delta h_{\rho\sigma})
 \\
 + \tilde{b} R^{\omega\lambda} (\partial^\rho \delta h_{\rho\lambda} - \partial_\lambda \delta h)
 + 2 \tilde{c} R (\partial_\rho \delta h^{\rho\omega} - \partial^\omega \delta h) . } \label{needgauge}
\end{equation}
 
If we simply use the canonical transformation $\delta h_{\rho\sigma} = - \partial_\beta h_{\rho\sigma} \delta x^\beta$ we will not have a gauge invariant expression, as in the case of the canonical Noether energy-momentum tensor in electrodynamics. Since we have an explicitly gauge invariant Lagrangian density, the Bessel-Hagen method \cite{besselhagen1921,jackiw1978} can be used to derive transformations of the $h_{\rho \sigma}$ which leaves the second term in the conservation law gauge invariant. This procedure involves solving for the field transformations such that coordinate invariance and gauge invariance is simultaneously preserved. The field transformation $\delta h_{\rho\sigma} = - \partial_\beta h_{\rho\sigma} \delta x^\beta + \partial_\rho \xi_\sigma + \partial_\sigma \xi_\rho$ with the most general vector $\xi_\sigma = \tilde{A} h_{\sigma\beta} \delta x^\beta + \tilde{B} h \eta_{\sigma\nu} \delta x^\nu$ must be substituted into (\ref{needgauge}) and solved for the parameters which preserve gauge invariance. We have a unique gauge invariant solution for $\tilde{A} = 1$, $\tilde{B} = 0$. Remarkably the solution is that this transformation is exactly $\delta h_{\rho\sigma} = - 2 \Gamma^\nu_{\ \rho \sigma} \delta x_\nu$, where $\Gamma^\nu_{\ \rho \sigma} = \frac{1}{2}(\partial^\nu h_{\rho\sigma} - \partial_\rho h^\nu_\sigma - \partial_\sigma h^\nu_\rho)$ is the linearized Christoffel symbol. For the second term in the conservation law we now have,

  \begin{equation}
  \fl
 \frac{\partial \mathcal{L}}{\partial (\partial_\omega \partial_\lambda h_{\rho\sigma})}\partial_\lambda \delta h_{\rho\sigma} =
( - 4 \tilde{a} R^{\omega\rho\lambda\sigma} R^\nu_{\ \rho\lambda\sigma}
 - 2 \tilde{b} R_{\rho\sigma} R^{\omega \rho \nu \sigma}
- 2 \tilde{b} R^{\omega\lambda} R^\nu_{\ \lambda}
 - 4 \tilde{c} R R^{\nu\omega}) \delta x_\nu . 
\end{equation}

The expression in brackets is manifestly gauge invariant without restriction of the free parameters $\tilde{a},\tilde{b},\tilde{c}$. We now consider the third term in (\ref{ourgenen}) with the transformation $\delta h_{\rho\sigma} = - 2 \Gamma^\nu_{\ \rho \sigma} \delta x_\nu$,

 \begin{equation}
 \fl
\eqalign{
\left( \partial_\lambda \frac{\partial \mathcal{L}}{\partial (\partial_\omega \partial_\lambda h_{\rho\sigma})} \right) \delta h_{\rho\sigma} =
\frac{1}{2} \left(  4 \tilde{c} + \tilde{b} \right) \left[  2 \partial^\omega R \eta^{\rho\sigma} - \partial^\sigma R \eta^{\rho\omega} - \partial^\rho \eta^{\sigma\omega}  \right] \Gamma^\nu_{\ \rho \sigma} \delta x_\nu
\\
+ \left( 4 \tilde{a} + \tilde{b}\right) \left[ 2 \partial^\omega R^{\rho\sigma} -  \partial^\sigma R^{\rho\omega} -  \partial^\rho R^{\sigma\omega} \right] \Gamma^\nu_{\ \rho \sigma} \delta x_\nu
 . } \label{thirdterm}
\end{equation}

The energy-momentum tensor can only be made gauge invariant with respect to the transformation $h_{\mu\nu}' = h_{\mu\nu} + \partial_\mu \xi_\nu + \partial_\nu \xi_\mu$ if (\ref{thirdterm}) vanishes because second order derivatives of the potential are required for an expression invariant under this transformation \cite{magnano2002}.  The only way for (\ref{thirdterm}) to vanish is therefore to fix the free coefficients of the Lagrangian density. There is a solution such that (\ref{thirdterm}) is identically zero (leaving the energy-momentum tensor in lowest integer), $\tilde{a} = \frac{1}{4}, \tilde{b} = -1, \tilde{c} = \frac{1}{4}$,

\begin{equation}
\mathcal{L} = \frac{1}{4} (R_{\mu\nu\alpha\beta} R^{\mu\nu\alpha\beta} - 4 R_{\mu\nu} R^{\mu\nu} + R^2) , \label{lagun}
\end{equation}

which is the linearized form of the Gauss-Bonnet Lagrangian! It was first discovered by Cornelius Lanczos in 1938 \cite{lanczos1938}, and has subsequently been a point of interest in many areas of both physics and mathematics \cite{escalante2004,charmousis2002,wheeler1986,granda2012,haghani2015,marugame2016}. From Noether's theorem a symmetric, conserved and gauge invariant energy-momentum tensor is derived,

\begin{equation}
\fl
T^{\omega\nu} = -  R^{\omega\rho\lambda\sigma} R^\nu_{\ \rho\lambda\sigma}
 + 2  R_{\rho\sigma} R^{\omega \rho \nu \sigma}
+ 2  R^{\omega\lambda} R^\nu_{\ \lambda}
 -  R R^{\nu\omega} + \frac{1}{4} \eta^{\omega\nu}(R_{\mu\lambda\alpha\beta} R^{\mu\lambda\alpha\beta} - 4 R_{\mu\nu} R^{\mu\nu} + R^2) . \label{gaussemt}
\end{equation}

The Belinfante procedure for higher order gravity can also be used to obtain this energy-momentum tensor \cite{belinfante1940, jackiw1994}, as well as the Fock method for deriving an energy-momentum tensor which is symmetric and conserved on shell \cite{fock2015,bivcak2016}. From both methods the resulting energy-momentum tensor is exactly what we have in (\ref{gaussemt}). 

Just like in classical electrodynamics the procedure for $N=2,M=2$ allows for the derivation of a completely gauge invariant model, where the energy-momentum tensor is symmetric, conserved and gauge invariant. The energy-momentum tensor presented in (\ref{gaussemt}) is a well known expression to string theorists for several decades \cite{ray1978,boulware1985,myers1987}. The fact that it follows from a procedure originally developed for deriving completely gauge invariant models in relativistic field theories (such as electrodynamics) was a completely unforeseen result. The connection between electrodynamics and Gauss-Bonnet gravity models, as well as the meaning of this connection, is worth further investigation.

\section{Conclusions}

A procedure was developed for building completely gauge invariant models by imposing gauge invariance and Noether's theorem to general scalar Lagrangian densities. The electrodynamic Lagrangian density is shown to that follows from this procedure for $N = 1$ and $M = 1$, which leads directly to the completely gauge invariant theory. For spin-2 ($N = 1$, $M = 2$), this procedure yields no nontrivial results, which is expected because the spin-2 Lagrangian density is only invariant up to a surface term. A model with $M = 2$ rank of potential (second rank symmetric potential $h_{\mu\nu}$) and $N = 2$ order of derivatives was derived from the procedure, yielding 3 possible contractions: linearized Riemann tensors, Ricci tensors, and Ricci scalars. It is found that for a specific combination of these terms, a completely gauge invariant model can be constructed analogous to electrodynamics; a model which is exactly the linearized Gauss-Bonnet gravity model.

If a Lagrangian density is built from a tensor potential of rank $M$ and order of derivatives $N = M$, the procedure can be used to derive a completely gauge invariant model; the Lagrangian density, equation of motion and energy-momentum tensor are all gauge invariant. This was highlighted by the fact that for $N = M = 1$ the procedure yields classical electrodynamics, and for $N = M = 2$ with a totally symmetric tensor potential the procedure yields linearized Gauss-Bonnet gravity. We note that this pattern continues for totally symmetric tensor potentials when $N = M > 2$ and is the subject of future work. In cases where $N \neq M$ it is possible to derive completely gauge invariant models if $N > M$, or for totally antisymmetric tensor potentials of any rank $M$ when $N = 1$ (keeping in mind the rank $M$ of totally antisymmetric potentials is restricted by the dimension $D$ of a theory as $M < D$).

A major characteristic highlighted by the developed procedure is the importance of a Lagrangian that is exactly gauge invariant in physical field theories; not simply invariant up to some surface term. This is explicit, but rarely discussed, in the from of the canonical energy-momentum tensor. The first term in (\ref{ourgenen}) must be independently gauge invariant, otherwise the only possible gauge invariant energy momentum tensor will be $T^{\lambda\gamma} = 0$. Electrodynamics ($N = 1$, $M = 1$) is a completely gauge invariant field theory built from a unique gauge invariant Lagrangian density. The gauge invariant Lagrangian for the $N = 2$ and $M = 2$ model presented in this article not only has this attribute, but it is directly related to the building blocks of general relativity through linearized Riemann tensors that were derived. Uniqueness found in the gauge invariant Lagrangian density expressions suggests that it is not enough to consider only the gauge invariance of the equation of motion. Complete gauge invariance of these models emphasizes the need for both exactly gauge invariant Lagrangians and conservation laws. Gauge invariant Lagrangian densities produced by this procedure imply existence of gauge invariant field strength tensors that can be used to build the corresponding model. The connection between Gauss-Bonnet gravity and classical electrodynamics is the primary point of interest presented in this article and will be the subject of future work. 

\section{Acknowledgement}

The authors are grateful to N. Kiriushcheva and D.G.C. McKeon for numerous discussions and suggestions during the preparation of our paper.

\section{Bibliography}

\bibliographystyle{unsrt}
\bibliography{MScThesisPaperBibliography}

\end{document}